\documentstyle[amssymb,preprint,prl,aps,12pt]{revtex}

\begin{document}

\begin{center}
\smallskip

{\bf Observation of Mammalian Similarity through Allometric Scaling Laws}

.

Valery B. Kokshenev \bigskip

{\it Departamento de F\'{i}sica, Universidade Federal de Minas Gerais, ICEx, 
}

{\it Caixa Postal 702, CEP 30123-970, Belo Horizonte, MG, Brazil \medskip }

(Submitted 9 September 2002, accepted 1 December 2002)

.

.

ABSTRACT

.
\end{center}

We discuss the problem of observation of natural similarity in skeletal
evolution of terrestrial mammals. Analysis is given by means of testing of
the power scaling laws established in long bone allometry, which describe
development of bones (of length $L$ and diameter $D$) with body mass in
terms of the growth exponents, {\it e.g.} $\lambda =d\log L/d\log D$. The
bone-size evolution scenario given three decades ago by McMahon was quiet
explicit on the geometrical-shape and mechanical-force constraints that
predicted $\lambda =2/3$. This remains too far from the mammalian allometric
exponent $\lambda ^{(\exp )}=0.80\pm 0.2$, recently revised by Christiansen,
that is a chief puzzle in long bone allometry. We give therefore new
insights into McMahon's constraints and report on the first observation of
the critical-elastic-force, bending-deformation, muscle-induced mechanism
that underlies the allometric law with estimated $\lambda =0.80\pm 0.3$.
This mechanism governs the bone-size evolution with avoiding skeletal
fracture caused by muscle-induced peak stresses and is expected to be unique
for small and large mammals.

.


.

{\bf PACs numbers:} 87.10.+e, 87.19.Ff, 87.23.Kg.\newpage

\smallskip {\bf Observation of Mammalian Similarity through Allometric
Scaling Laws}

\begin{center}
by V.B. Kokshenev\bigskip
\end{center}

\section{Introduction}

In general, biological laws do not follow from physical laws in a simple and
direct way. Examples include Kleiber's allometric law known as the $3/4$
power law that scales metabolic rates for animals and plants to their mass
within the range of three order of magnitude. As shown by West {\it et al}.
in Refs.\cite{WBE97,WBE99} the observed metabolic rate scaling law arises
from the interplay between geometric and physical constraints implicit,
respectively, in space-filling fractal networks and energy dissipation%
\footnote{%
For simple explanations of Kleiber's allometric scaling, which is shown to
originate from the general features of the networks irrespective of the
geometrical and dynamical details, see Refs.\cite{BMR99,BMR00,Dr01}}.
Another famous $2/3$ power law was proposed by McMahon\cite{Mc73} for
scaling of longitudinal-to-transverse dimensions of animals and plants
through physical description of geometric-shape and critical-force
similarities noticed in their size evolution. Given in Ref.\cite{Mc73} in
explicit form, the geometrical-(cylindrical-volume)-shape and
mechanical-(critical-elastic-buckling)-force constraints imposed on size
evolution for animals and plants with their mass yielded the $2/3$ power
scaling law, along with the $1/4$ and $3/8$ laws deduced\cite{Mc73},
respectively, for longitudinal and transverse linear dimensions. During
almost three decades, McMahon's scaling laws have been a controversial
subject of intensive study and debate. As a matter of fact, McMahon's
description of the geometrical-shape and mechanical-force similarities was
experimentally proved for terrestrial mammals neither in body allometry\cite
{E83} nor in long bone allometry\cite
{Ale77,AJM79,AJM81,B83,Hok86,BB92,C99a,C99b,C02}. Moreover, the most recent
condemnation by Christiansen\cite{C99b} states that no satisfactory
explanation for any power-law scaling observed in mammalian allometry can be
expected.

We will demonstrate that the failure of McMahon's constraints is due to the
fact that the skeletal subsystem of animals is not mechanically isolated
from their muscle subsystem, as was suggested in Ref.\cite{Mc73}. Also,
McMahon's hypothesis that the skeletal support of weight and fast locomotion
of mammals is driven solely by a gravitation contradicts to up-to-date
comprehension on a role of muscle fibers and tendons in formation of maximum
skeletal stresses. We therefore revisit McMahon's evolution constraint
equations in Sec.II within the context of their application to long bone
allometry for terrestrial mammals. These equations are modified and
generalized in view of the known experimental findings in muscle fiber
allometry. Experimental testing of the two distinct critical-elastic-force
mechanisms that govern evolution of mammalian bones is elaborated in Sec.
III. Discussion and conclusions are given in Sec. IV.

\section{ McMahon's Constraints in Long Bone Allometry}

\subsection{Elastic Similarity Model Revisited}

Famous power laws by McMahon\cite{Mc73} for scaling of linear dimensions of
animals and plants was proposed within the framework of the so-called
elastic similarity model (hereafter, ESM). Application of the ESM by McMahon
to the case of mammalian bone allometry was based on the {\em cylindric-shape%
} correspondence that takes place between a given skeletal bone and a
cylindrical beam. A bone sample was therefore geometrically approximated by
a cylinder of diameter $D_{is}$ and length $L_{is}$, where index $i$ counts
different bones and $s$ indicates mammalian specie. The {\em mechanical-force%
} correspondence to the same {\em rigid} cylinder is justified by
observation of the universal (specie-independent) bone-stress safety
factors. These are given by ratio (about $3$) of yield stress to peak
stress. Exploration of such a kind of mechanical correspondence by McMahon
gave rise to the maximum-(elastic-buckling)-force {\em constraint} imposed
on volume-size evolution of a given bone.

More specifically, the ESM is based on the fact that the mechanical failure
of a bone is prevented through its linear dimensions $D_{is}$ and $L_{is}$,
adjusted to bear critical {\em buckling} {\em deformations}, related to peak
stresses through the maximum elastic forces: $F_{elast}^{(\max
)}=F_{buckl}^{(crit)}$. The latter is describes an elastic instability
caused by critical bending deformations that was specified by the Euler
critical estimate $F_{buckl}^{(crit)}={\pi ^{2}EI/L^{2}}$ for a given
cylinder (of length $L$ and of diameter $D$, with the moment of inertia $%
I=\pi D^{4}/64$ and the elastic modulus $E$, see {\it e.g.} Cap.IV in Ref.%
\cite{LaL}). Thus the ESM constraint equations attributed by McMahon to the
cylindric-shape and elastic-force similar skeletal bones can be introduced
through ($a$) the elastic-buckling critical force $F_{is}^{(crit)}$ and ($b$%
) the cylindric-bone volume $V_{is}^{(bone)}$, namely

\begin{eqnarray}
\text{ }F_{is}^{(crit)} &=&\frac{\pi }{64}E\frac{D_{is}^{4}}{L_{is}^{2}} 
\eqnum{1a} \\
V_{is} &=&D_{is}^{2}L_{is}\text{ .}  \eqnum{1b}
\end{eqnarray}

In long-bone allometry, the observation of evolution of limb bones across
mammalian species is discussed though the bone-size linear-dimension scaling
to body mass $M_{s}$. This is given in terms of the bone-diameter and the
bone-length allometric exponents, respectively, $d_{i}$ and $l_{i}$, or of
the $i$-bone-dimension{\em \ growth exponents}, introduced by the following
scaling differential relations, namely 
\begin{eqnarray}
\text{ }d_{i} &=&\frac{d\log D_{is}}{d\log M_{s}}\text{, }l_{i}=\frac{d\log
L_{is}}{d\log M_{s}},\text{ and}  \eqnum{2} \\
\lambda _{i} &=&\frac{d\log L_{is}}{d\log D_{is}}\equiv \frac{l_{i}}{d_{i}}.
\eqnum{3}
\end{eqnarray}
The {\em reduced dimension} exponent $\lambda _{i}$, related to the
longitudinal-to-transverse scaling, is also defined. As seen, Eqs.(2),(3)
are equivalent to the corresponding differential equations $%
dD_{is}/dM_{s}=d_{i}D_{is}/M_{s}$, {\it etc.}, which solutions are commonly
derived in bone allometry through regression equations $%
D_{is}=c_{is}M^{d_{i}}$, where $M$ is treated as an external mammalian
parameter and $c_{is}$ are constants. A notable feature of the introduced
scaling differential relations is independence of the $i$-bone exponents on
mammalian specie $s$. This corroborates the bone allometry observations and
Eqs.(2),(3) are therefore treated as the allometric {\em scaling laws}. This
implies a universal fashion in evolution of any linear dimension of bones,
as well as bone {\em volume }$V_{is}$ $=D_{is}^{2}L_{is}$, with body mass
that in a certain way reflects similarity of mammals with their size
evolution. With taking into account that $\rho V_{is}=M_{is}$ ($\rho $ is
bone density), and adopting additionally McMahon's hypotheses that ($a$)
effective skeletal growth is driven by gravitation, {\it i.e.}, $%
F_{is}^{(crit)}$ ${\thicksim }$ $gM_{is}$ ($g$ is the gravity constant) and
that ($b$) bone mass $M_{is}$ linearly scales to body mass $M_{s}$, the
following $i$-bone-evolution equations, namely{\em \ }

\begin{equation}
{\em \ }\left\{ 
\begin{array}{c}
4d_{i}-2l_{i}=1, \\ 
2d_{i}+l_{i}=1
\end{array}
\right.  \eqnum{4}
\end{equation}
result from, respectively, Eqs.(1a) and (1b) with the help of the scaling
relations given in Eq.(2). In turn, Eqs.(4) and (3) provide the well known
ESM predictions: $d_{0}^{(buckl)}=3/8$, $l_{0}^{(buckl)}=1/4$, and $\lambda
_{0}^{(buckl)}=2/3$, including a trivial {\em isometric }solution{\em \ }$%
d_{0}=l_{0}=1/3$ and $\lambda _{0}=1$. As mentioned in the Introduction,
these predictions were not experimentally proved even when a statistical
dispersion of allometric data was taken into account (for recent criticism
of the ESM predictions for the allometric exponents $d,$ $l$ and $\lambda $
see analyses given in Table 5 in Refs.\cite{C99a} and Table 3 in Ref.\cite
{C99b}, respectively).

\subsection{Elastic-Buckling-Force Criterium}

Skeletal evolution of animals cannot be studied independently of their
muscle fibers and tendons. Moreover, the peak skeletal stresses are
generated rather by muscle contractions than by gravitation. These both
statements follow from studies of muscle design and bone strains during
locomotion\cite{CSS80,RL84,SC89,B91}. We infer therefore that the maximum
elastic forces exerted by long bones are originated from the {\em maximum}
muscle forces, {\it i.e.}, $F_{elast}^{(\max )}=F_{musc}^{(\max )}$. The
same studies provide strong evidence that the maximum muscle stresses are
independent of body mass, and thus $F_{musc}^{(\max )}/A_{musc}^{(\max
)}\propto M^{0}$, where $A_{musc}^{(\max )}$ is the maximum cross-section
area of muscle fibers. The critical-force constraint, justified through the
aforementioned bone-stress safety factors, can be therefore formally
introduced into consideration by the ''overall-bone'' {\em critical-force
exponent} $a_{c}$, namely 
\begin{equation}
a_{c}=\frac{d\log F_{musc}^{(crit)}}{d\log M_{s}}\text{ }=\text{ }a_{cm}=%
\frac{d\log A_{musc}^{(\max )}}{d\log M}  \eqnum{5}
\end{equation}
and the corresponding {\em critical muscle-area} exponent $a_{cm}$. These
should be distinguished from the exponents 
\begin{equation}
a_{ci}=\frac{d\log F_{is}^{(crit)}}{d\log M_{s}}\text{ and }a_{m}=\frac{%
d\log A_{musc}}{d\log M}  \eqnum{6}
\end{equation}
where $F_{is}^{(crit)}$ is given in Eq.(1a). The muscle-area exponent $a_{m}$
is known in muscle allometry\cite{AJM81,SC89,PS94} as the muscle-fiber,
cross-section-area exponent and can be exemplified by data $a_{m}^{(\exp
)}=0.69-0.91$ derived by Pollock and Shadwick for four distinct groups of
muscles in mammalian hindlimbs (see Fig.3 in Ref.\cite{PS94}). The maximum
muscle force is commonly associated\cite{AJM81} with the {\em leg group} of
muscles of animals, {\it i.e.}, $A_{musc}^{(\max )}=A_{musc}^{(leg)}$. With
adopting of the latter in Eq.(5), the ''leg-muscle'' {\em critical} exponent 
$a_{cm}^{(\exp )}=0.81-0.83$ was obtained\cite{PS94} (on the bases of data%
\cite{AJM81} for six groups of mammalian leg muscles by Alexander {\it et al.%
}) and reported by Pollock and Shadwick in Ref.\cite{PS94}. As seen, the
means $\overline{a}_{m}^{(\exp )}=0.80$ and $\overline{a}_{cm}^{(\exp
)}=0.82 ${\it \ }are different but not distinguished within the experimental
error.

In order to establish the critical muscle-area exponent $a_{cm}^{(\exp )}$
defined in Eq.(5), we have reanalyzed the experimental data by Pollock and
Shadwick on muscle fiber area $A_{musc}$ in hindlimbs of 35 quadrupedal
mammalians as a function of body mass represented in {\bf Fig.1} from Fig.3
in Ref.\cite{PS94}. In general, the {\em gastrocnemius} group of muscles
(shown by diamonds in Fig.1 for points $A_{m}^{(G)}$ adjusted\cite{PS94}
with $a_{m}^{(G)}=0.77$), unlike the {\em common digital extensors} group
(shown by crosses for points $A_{m}^{(C)}$ with\cite{PS94} $a_{m}^{(C)}=0.69$%
), plays a principal role in formation of maximum bone stresses. This is due
to the fact that for small and large species of animals $%
A_{m}^{(G)}>A_{m}^{(D)}$. Other two groups of muscles exhibit a crossover
from the almost isotropic evolution with $a_{0}=2/3$ to somewhat given by
the exponent $a_{cm}$ and driven by the maximum muscle areas controlled by
the {\em gastrocnemius} group, {\it i.e.} by $A_{cm}$ established by maximum
points of $A_{musc}^{(G)}$. First the highest 5 points (shown by arrows in
Fig.1) are fitted by $A_{cm}^{(1)}=428*M^{0.80}$. Next nearest-neighbors (16
points indicated as solid symbols below the line $A_{cm}$ in Fig.1) are
fitted with $A_{cm}^{(2)}=304*M^{0.83}$. Regression elaborated within all
the field of a maximum muscle area, defined by the highest 21points,
provides $a_{cm}^{(1,2)}=0.82\pm 0.01$ (with correlation coefficient $%
r=0.996 $). Remarkably, this finding matches well the aforegiven data for
the ''leg-muscle'' exponent by Alexander {\it et al.} reported in Ref.\cite
{PS94} and can be therefore treated as a reliable data.

Eqs.(5) and (6) provide the following definition for the ''overall-bone''
averaged exponents, namely 
\begin{equation}
a_{c}=<a_{ci}>\equiv \frac{1}{n}\sum_{i=1}^{n}a_{ci}=a_{cm}\text{, with }%
a_{cm}^{(\exp )}=0.82\pm 0.01,  \eqnum{7}
\end{equation}
where summation is limited by bones which do play a {\em principal} role in
effective support and fast locomotion of body mass of animals. Eq.(7) can be
also treated as an extension of a similar definition of the {\em mammalian}
principal-bone-averaged exponents $d$, $l$ and $\lambda $ introduced with
the help of Eqs.(2),(3), {\it e.g.}, $d=<d_{i}>$. Thereby, revision of
McMahon's $a$-hypothesis provides a new $a$-constraint equation imposed on
the exponents: $4d-2l=a_{c}$.

In view of the fact that neither skeletal mass\cite{Hok86} nor bone mass\cite
{C02} are linear with mammalian body mass, McMahon's revised $b$-constraint
equation given in Eq.(4) for $i$-bone is also modified as $%
2d_{i}+l_{i}=b_{i} $. Here the $i$-{\em bone}-{\em mass} exponent, namely

\begin{equation}
b_{i}=\frac{d\log M_{is}}{d\log M_{s}}\text{.}  \eqnum{8}
\end{equation}
is introduced through the relevant power scaling law. Thus, McMahon's
critical-force and cylindric-shape constraints given in Eq.(4) result in the
following {\em modified} ESM constraints, namely

\begin{equation}
\left\{ 
\begin{array}{c}
4d-2l=a_{c}, \\ 
2d+l=b.
\end{array}
\right.  \eqnum{9}
\end{equation}
In turn, this yields new predictions for the mammalian overall-bone
dimension and reduced-dimension exponents, or the {\em %
elastic-buckling-criterium} predictions, namely 
\begin{eqnarray}
d^{(buckl)} &=&\frac{a_{c}+2b}{8}\text{ , }l^{(buckl)}=\frac{2b-a_{c}}{4}%
\text{ and }  \eqnum{10} \\
\lambda ^{(buckl)} &=&8<\frac{b_{i}}{a_{ci}+2b_{i}}>-2\text{ .}  \eqnum{11}
\end{eqnarray}
The latter prediction follows from the definition for the reduced-dimension
exponent $\lambda _{i}=l_{i}/d_{i}$ given in Eq.(3) and presented here in
the form $\lambda _{i}=b_{i}/d_{i}-2$, with the help of the $b$-constraint
equation.

\subsection{Elastic-Bending-Force Criterium}

After Alexander {\it et al}.\cite{AMH79} it has been widely recognized (for
recent references see Ref.\cite{C02}) that the elastic {\em bending}
deformations play a crucial role in the overall peak stresses of long bones
instead of a simple axial compression discussed\cite{Mc73} by McMahon in
terms of the critical buckling deformations. The corresponding critical
force $F_{elas}^{(\max )}=F_{bend}^{(crit)}$ applied normally to the bone
before fracture was already discussed in long-bone allometry in Refs.\cite
{Hok86,SC89}. In view of the elastic nature common for both kind of
deformations, the force $F_{bend}^{(crit)}$ in a certain way extends the ESM
given in Eq.(1) for the case of the bending critical deformations, namely

\begin{eqnarray}
F_{is}^{(crit)} &\thicksim &E\frac{D_{is}^{3}}{L_{is}},  \eqnum{12a} \\
\rho D_{is}^{2}L_{is} &=&M_{is}.  \eqnum{12b}
\end{eqnarray}
Straightforward application of the scaling differential relations introduced
in Eqs.(2),(3), with accounting of the critical-force and the bone-mass
growth exponents given in, respectively, Eqs.(5),(8) and (9), results in the
following new constraint equations:

\begin{equation}
\left\{ 
\begin{array}{c}
3d-l=a_{c}, \\ 
2d+l=b.
\end{array}
\right.  \eqnum{13}
\end{equation}
This provides the {\em elastic-bending criterium} expressed in terms of the
following predictions for the mammalian bone-dimension growth exponents,
namely 
\begin{eqnarray}
d^{(bend)} &=&\frac{a_{c}+b}{5}\text{ , }l^{(bend)}=\frac{3b-2a_{c}}{5}\text{
and }  \eqnum{14} \\
\lambda ^{(bend)} &=&5<\frac{b_{i}}{a_{ci}+b_{i}}>-2.  \eqnum{15}
\end{eqnarray}
Notably that both the elastic-force criteria given in Eqs.(10) and (14) are
consistent with the isometric solution ($d_{0}=l_{0}=1/3$ and $\lambda
_{0}=1 $), which is found under conditions that the mammalian muscle-area
subsystem develops isometrically ($a_{0}=2/3$) and independently of the
skeletal subsystem ($b_{0}=1$). The observed allometric scaling laws with $%
d^{(\exp )}>0.33$, $l^{(\exp )}<0.33$, and $\lambda ^{(\exp )}<1$
corroborate that this simplified geometric scenario is avoided by the nature.

\section{Observation of Bone Evolution Similarities through Experimental
Testing of the Constraint Equations}

All predictions given by the original ESM\cite{Mc73} and the revised ESM are
analyzed in the bone growth diagram in {\bf Fig.2}. As seen, the available
experimental data matches neither the isometric nor the original ESM
solutions (shown by crosses), even in case when dispersion effects of the
experimental data (shown by error bars) are taken into account. Note that
this large dispersion is not caused by error measurements of bone dimensions
or body mass of animals, but is resulted from a large phylogenetic spectrum
of terrestrial mammals\footnote{%
In fact, there exist a certain error due to deviation of bone shape from the
ideal cylinder. Also, not all body mass were really measured but taken as an
average from the literature data (see discussion in Ref.\cite{C99b}).}.
Unlike the case of the pioneer data\cite{AJM79} by Alexander {\it et al.},
all species which have multiple specimens, were additionally averaged\cite
{C99b} within a certain mammalian subfamily before to be documented. The
most accurate allometric data with the systematically reduced phylogenetic
statistical error was given\cite{C99a,C99b,C02} by Christiansen.

Predictions of the modified ESM and the extended ESM are shown in Fig.2 by
the shaded areas, which correspond, respectively, to Eqs.(9) and (13)
estimated with account of the reliable domain for the critical-force
exponent $a_{c}^{(\exp )}=0.81-0.83$ and of that for the bone-mass exponent $%
b^{(\exp )}=1.0-1.1$ (that approximately covers error scatter of the
experimental data on $b_{i}^{(\exp )}$(given in Table 2 in Ref.\cite{C02}).
The shaded areas indicate the critical-force constraints given by the $a$%
-constraint lines $4d-2l=0.82$ and $3d-l=0.82$ extended by cylindric-volume
constraints implicit in the form of the elastic-buckling-force and
elastic-bending-force criteria, respectively. As seen from Fig.2, the
elastic-{\em buckling} criterium seems to be observable within the range of
the unreduced phylogenetic statistical error. After reduction of this error,
only the elastic-{\em bending} criterium corroborates.

Besides the case of the 6-long-bone-averaged allometric data\cite{C99a}
given in Fig.2 for the {\em one-scale} least-square regression ({\em LSR}),
we have also elaborated analysis of the double set of the allometric
exponents (taken from Table 5 in Ref.\cite{C99a}) derived within the {\em %
two-scale} regressions made for small ($M<50kg$) and large ($M>50kg$)
mammals. But no definitive conclusions on domination of any elastic-force
criteria can be inferred. Indeed, in the case of the overall-($6$
-bone)-average analysis, the data for small and large animals is far to be
fitted by any of the dashed areas in Fig.2. When the {\em ulna} and the {\em %
fibula} are excluded, the principal-($4$-bone)-averaged {\em LSR} data
justifies the elastic-bending and the elastic-buckling criteria for the
cases of small and large mammals, respectively. However, unlike the case of
the one-scale data, experimental accuracy of the two-scale analysis is
marginal that makes doubtful any inference on observation of both the
distinct critical-force constraints. We have therefore restricted our
analysis by one-scale allometric data for the four principal mammalian long
bones listed in {\bf Table 1}.

First, we check a {\em self-consistency} of experimental data on the
dimension ( $l_{i}^{(\exp )}$, $d_{i}^{(\exp )}$ )\cite{C99a} and
reduced-dimension ($\lambda _{i}^{(\exp )}$ )\cite{C99b} allometric
exponents obtained independently and presented in first and second columns
of Table 1, respectively. As seen, when the bone-averaged data is compared
between the two regression methods, it obeys the relation $d^{(\exp )}$/$%
l^{(\exp )}=$ $\lambda ^{(\exp )}$ with accuracy that is much higher than
that for the case of partial $i$-bone relations $d_{i}^{(\exp )}$/$%
l_{i}^{(\exp )}=$ $\lambda _{i}^{(\exp )}$ compared within the same method.
Then, the geometrical mammalian similarity is tested on the basis of the $b$%
-constraint equation $2d_{i}^{(\exp )}+l_{i}^{(\exp )}=b_{i}^{(\exp )}$ in
second and third columns of Table 1. Again, the cylindric-shape constraints,
given in terms of the bone-averaged data, are justified\footnote{%
Exclusion should be given for the case of the exponent $b^{*}$ , which data
obtained by the square regression (LSR) method is not available.} with a
good precision. We infer that observation of the mammalian similarity
through the allometric power laws can be realized only ''on the average'',
but not for a given type of ''mammalian'' bone as it widely adopted in
allometric studies. Examples are analyses of the original ESM predictions
elaborated for a given $i$-bone, instead of the overall-bone data, and given
in Table 5 in Ref.\cite{C99a}, Table 3 in Ref.\cite{C99b}, and Table 3.11 in
Ref.\cite{Gar01}.

The problem of validation of the bone-evolution $a$-constraint equation for
the case of the bending loads, {\it i.e.}, $3d-l=a$, where $a$ is treated as
a free parameter, was first discussed\cite{SC89} by Selker and Carter in
terms of the bone strength index. On the basis of the mammalian data\cite
{B83} by Biewener (shown in Fig.2) and their own data for {\em artiodactyls}%
, the overall-bone-averaged equation $3d^{(\exp )}-l^{(\exp )}=a$ provided%
\cite{SC89} estimates $a=0.77$ and $0.82$, respectively. By generalization
of these findings to the overall mammalian case, allometric exponent $%
a_{m}^{(\exp )}=0.77-0.82$ was adopted\cite{SC89} for testing of the bending
or torsion deformations in mammalian long bones due to muscle contractions.
The same analysis made on the basis of other available in the biological
literature allometric data, including the particular case of birds\footnote{%
Application of the ESM for birds remains questionable.}, has been recently
given by Garcia (see Table 3.12 in Ref.\cite{Gar01}). As the result, the
allometric muscle-area exponent was suggested $a_{m}^{(\exp )}=0.77-0.83$,
with the mean $\overline{a}_{m}^{(\exp )}=0.80$, as a suitable data for
experimental testing of the bending-force constraint equation (see analysis
in Table 3.11 in Ref.\cite{Gar01}). This suggestion is not true.

Indeed, as follows from the pioneer work\cite{Mc73} by McMahon, revisited in
the previous section, the force-constraint equation is driven by the {\em %
critical} force and therefore given as $3d-l=a_{c}$ where the critical-force
exponent, according to Eqs.(5),(7), is established by the data on
maximum-muscle-area allometry, {\it i.e.}, $a_{c}=a_{cm}^{(\exp )}=0.81-0.83$%
, with $\overline{a}_{cm}^{(\exp )}=0.82$ that should be distinguished from
the suggested\cite{Gar01} data on $\overline{a}_{m}^{(\exp )}=0.80$. We have
therefore reconsidered analysis given in Table 3.11 in Ref.\cite{Gar01} and
found\cite{tobe} that no definite conclusions can be made on validation%
\footnote{%
Again, the marginal estimate $a_{cm}=0.829$ is obtained in the case of the
small-animal {\em LSR} data.} on the principal-bone averaged equation $%
3d^{(\exp )}-l^{(\exp )}=a_{cm}^{(\exp )}$ on the bases of the {\em two-scale%
}{\it \ RMA}\ and {\it LSR} data\cite{C99a}. Conversely, the {\em critical}%
-bending-force constraint equation $3d-l=a_{c}$ is strongly supported by the 
{\em one-scale} data\cite{C99a} by Christiansen deduced through both the
different ({\it LSR} and {\it RMA}) regressions. This follows from the
bone-dimension predictions $a_{c}=0.82$ and $0.83$ (obtained\cite{tobe},
respectively, for both the methods with the help of data given in first
column in Table 1).

In the current study we put emphasis on observation of the mammalian
similarity through the critical muscle allometry exponent $a_{cm}^{(\exp )}$
established in Fig.1 along with the one-scale long-bone allometric data on
the reduced-dimension exponent $\lambda _{i}^{(\exp )}$obtained in Ref.\cite
{C99b}. Therefore, we have reformulated the elastic-buckling and the
elastic-bending criteria given in Eqs.(11) and (15) in terms of the
observable $\lambda _{i}$ . This provides the following predictions for the
critical-force exponents, namely

\begin{equation}
a_{c}^{(buckl)}=2<\frac{2-\lambda _{i}}{2+\lambda _{i}}b_{i}>\text{ and }%
a_{c}^{(bend)}=<\frac{3-\lambda _{i}}{2+\lambda _{i}}b_{i}>  \eqnum{16}
\end{equation}
obtained with the help of Eq.(7) and estimated in last column of Fig.1. As
seen, the elastic-force bone-{\em buckling}-deformation mechanism, proposed
by McMahon in Ref.\cite{Mc73} suggests an estimate $a_{c}^{(buckl)}=0.90$
for both the regression methods that is not justified by Eq.(7). In
contrast, the elastic-force bone-{\em bending}-deformation mechanism
predicted by $a_{c}^{(bend)}=0.81$ (and $0.83$) within the {\em LSR} (and 
{\em RMA} regression) methods is proved by the reduced-dimension and
self-consistent linear-dimension allometry data reported by Christiansen in
Refs.\cite{C99b} and Ref.\cite{C99a}, respectively. Again, analysis, similar
to that given in Table 1, extended to the case of the {\em two-scale}
principal-bone data\cite{C99a,C99b}, corroborates the same bending-force
mechanism only for the data derived for {\em small} mammals within the {\it %
LSR} method. No conclusions can be inferred in the case of small mammals
treated by the {\em RMA} regression, as well as in the case of large mammals
treated by both the methods.

\section{Discussion and Conclusions}

We have discussed the problem of observation of natural similarity in
evolution of terrestrial mammals. Testing of the two conceivable underlying
mechanisms that drive the bone size development with body mass of animals is
given on the basis of experimental data on the reduced dimension ($\lambda
_{i}=$ $l_{i}/$ $d_{i}$), longitudinal dimension ($l_{i}$), and transverse
dimension ($d_{i}$) allometric exponents established through the scaling
laws commonly discussed in the mammalian long-bone allometry.

Since Galilei it was repeatedly recognized that the isometric skeletal
evolution prescribed by the overall-bone exponent $\lambda _{0}=1$ is not
observed in the nature because the small mammals are not geometrically
overbuilt and the large species do not operate very close to their
mechanical failure limit, that it would be expected from the isometric
scenario. This figurative, widely cited description given by Biewener\cite
{Bie90} is in agreement with the simplified version ($\lambda _{0}=1$) of
the more sophisticated scenario ($\lambda _{0}^{(buckl)}=0.667$) proposed%
\cite{Mc73} by McMahon. Within the ESM, the mammalian similarity was
introduced\cite{Mc73} on the basis of realistic geometrical-shape and
mechanical-force correspondence that takes place between a given skeletal
bone and a rigid cylinder. As mentioned in the Introduction and illustrated
in Fig.2, predictions for evolution of bone dimensions with body mass given
by the ESM were disapproved in long bone allometry. This, in particular,
implies that the ESM exponent $\lambda _{0}^{(buckl)}=0.667$ was not
justified in long-bone allometric experiments, including the most systematic
data with $\lambda ^{(\exp )}=0.78-0.82$ (that follows from Table 1 as the
mean between the {\it LSR} and {\it RMA} bone-averaged data).

A good deal effort has been undertaking in long bone allometry to learn
experimental conditions for observation of the critical-force (elastic-{\em %
buckling}-deformation and {\em gravity}-induced) mechanism suggested by
McMahon for explanation of anatomical adaptation of skeletal bones through
their linear dimensions. The first objection\cite{E83} by Economos was as
follows. McMahon's mechanical-failure mechanism should not be expected as a
unique for all species, but would more suitable for {\em large} mammals.
This stimulated a careful search for additional scaling laws related to
small and large mammals. Such kind of new scaling laws were established in
terms of the double sets of allometric exponents introduced\cite
{C99a,C99b,C02} by Christiansen through the {\em two-scale} regressions
distinguished by $M_{c}=50kg$ adopted as a boundary mass common for all
species. Furthermore, it was speculated that the revealed inadequate
description of the scaling laws is due to inaccuracy of the methods of
regression and, as a result, the {\em RMA} regression was suggested\cite
{C99a} as well-chosen instead of the traditional {\em LSR}. The second
objection\cite{E83} by Economos refers to the {\em linearity} of the
logarithmic scaling laws given in Eq.(2), which was not expected to be a
sole across the three order of magnitudes of body mass. Experimental
verification of this idea by Christiansen revealed\cite{C99a} that an
application of the polynomial type of regressions in bone allometry does not
improve the correlations established within the traditional linear
logarithmic scaling. Finally, thorough numerical analyses\cite{C99a,C99b} of
the reasons of the ESM failure brought Christiansen to a conclusion that
''many factors contribute to maintaining skeletal stress at uniform level'',
including the factor of bending-deformation-induced stresses, which are more
important\cite{C99a} than the bone stresses illuminated\cite{Mc73} by
McMahon.

We have demonstrated how the factors of muscle fiber contractions, bone mass
evolution, and of bending bone deformations can be incorporated into the ESM
constraint equations. As a result, the {\em modified} (by bone-mass and
muscle-contraction factors) ESM becomes observable (see shaded area that
extends $a$-buckling line in Fig.2) under condition that the unreduced
phylogenetic statistical dispersion of the allometric data\cite
{AJM79,B83,BB92} is taken into account. Otherwise, the {\em extended }%
(additionally by the {\em bending}-deformation factor) ESM agrees with
experiment (see Fig.2). Another analysis (given in Table 1) yields the
observation of the mammalian similarity within the {\em principal}-long-bone
allometric data\cite{C99a,C99b,C02} with systematically reduced statistical
error by Christiansen. As demonstrated, this observation is realized in
terms of the bone-{\em averaged} allometric exponents, restricted by the 
{\em principal} bones that are involved into the evolution-constraint
equations. Example is the volume-constraint equation, which should be valid
for any conceivable bone-evolution mechanism. As seen from analysis given in
columns 2 and 3 in Table 1, the volume-constraint equation is much better
''observed'' in the ''bone-averaged'' form $2d^{(\exp )}+l^{(\exp
)}=b^{(\exp )}$ than in the form presented for a given $i$-bone. We guess
that the observation of the geometric-shape similarity through experimental
justification of exact Eq.(12b) should depend on neither the number of
scales nor on methods chosen for regression of the bone-dimension allometric
data. Our additional verification of the cylindric-shape similarity given on
the basis of the {\em two-scale} {\em principal}-long-bone allometric data
(taken from Table 3 in Ref.\cite{C99a}) corroborates this statement for both
small and large mammals. We infer therefore that both the methods and both
the scales are equivalent in observation of the ''bone-averaged''
geometrical-shape mammalian similarity, at least for the principal\footnote{%
Extended statistical analysis of both the constraint equations, with
including all available bone allometric data will be discussed elsewhere\cite
{tobe}.} bones.

As highlighted by Christiansen, the principal bones play a crucial role in
primarily support the body mass. He noted\cite{C99b} that greatly reduced 
{\em ulna} and too thin {\em fibula} do not play of much importance in
support of body mass. They are therefore not suitable for testing of the
critical-force constraints and should be excluded from the principal bone
set. As follows from our analyses given in Fig.1, qualitatively the same
should be referred to some muscle fiber groups such as {\em common} {\em %
digital extensors\ }which eventually do not produce peak bone stresses.
Meanwhile, as seen from Fig.9 in Ref.\cite{C99b} and Fig.1, {\em tibia} and 
{\em plantaris} show a crossover behavior between principal and
non-principal sets, where the principle bone-set and the principle
muscle-group set are presented by small and large mammals, respectively. As
the reliable {\em critical} principal-muscle-area exponent $a_{cm}$ ($=a_{c}$%
), which enters the critical-force $a$-equations, the data $a_{cm}^{(\exp
)}=0.82\pm 0.01$ is proposed in Eq.(7). This finding is derived in Fig.1
from the {\em maximum} muscle-area gastrocnemius group of leg muscles and
should be distinguished from the {\em overall} muscle-area data $%
a_{m}^{(\exp )}=0.80\pm 0.03$ that was groundlessly used, instead of $%
a_{cm}^{(\exp )}$, in establishing of experimental validation\cite
{SC89,Gar01} of the critical-{\em bending}-force constraint $3d^{(\exp
)}-l^{(\exp )}=a_{cm}^{(\exp )}$. As shown, this equation, unlike the case
of the critical-{\em buckling}-force constraint $4d^{(\exp )}-2l^{(\exp
)}=a_{cm}^{(\exp )}$ related to the original ESM, is observable directly and
indirectly through, respectively, the $a$-constraint equation and Eq.(16)
(analyzed in the last column of Table 1). Again, we infer that the
observation of the {\em bending}-force criterium does not depend on the
method chosen within the {\em one-scale} regression.

This is not the case for the{\em \ two-scale} data on bone-dimension
allometric exponents reported\cite{C99a} by Christiansen. Indeed, as follows
from our many-sided analysis, the elastic-bending criterium is definitely
supported for {\em small} and large mammals within the
(principal-bone-averaged) {\em LSR} data and {\em RMA}, respectively. With
accounting of the observation of the same criterium though the {\em one-scale%
} (principal-bone-averaged) {\em LSR} data, we see that correlations
established by the traditional {\em LSR }method, unlike suggestion given in
Ref.\cite{C99a}, show their self-consistency. But no certain conclusions can
be inferred within the observation windows for small and large mammals in
the cases of, respectively, {\em RMA} regression and {\em LSR}. We guess
that the revealed discrepancy of the two equal in rights regression methods
signals on failure of definition of the observation windows employed for the
analysis of the critical-force constraints. In other words, unlike the case
of the cylindric-shape similarity, these windows are not expected to be
universal for observation of the elastic-force mechanisms, and cannot be
therefore introduced by the unique boundary mass $M_{c}$.

Thereby we have demonstrated that the mammalian similarity, observable
through experimental validation of the bone-evolution constraint equations,
is described in terms of the {\em one-scale} {\em principal}-bone-{\em %
averaged} characteristics, which show independence on the regression
methods. Within this context, the observed in nature long-bone mammalian
evolution can be described through longitudinal-to-transverse bone-dimension
scaling law, with the aforegiven ''method-averaged'' exponent $\lambda
^{(\exp )}=0.80\pm 0.02$. Assuming a high enough accuracy for the $i$-bone
experimental data on the exponents $a_{ci}^{(\exp )}$ and $b_{i\text{ }%
}^{(\exp )}$, both the discussed evolution-mechanism criteria are
approximated in the following forms, namely 
\begin{equation}
\lambda ^{(buckl)}=2\frac{2b-a_{cm}}{2b+a_{cm}}\text{ and }\lambda ^{(bend)}=%
\frac{3b-2a_{cm}}{b+a_{cm}}  \eqnum{17}
\end{equation}
that follows from Eqs.(11) and (15), respectively. With accounting of $%
a_{cm}^{(\exp )}=0.81-0.83$, and adopting for the bone-mass mammalian
allometric exponent $b^{(\exp )}=1.03-1.06$ (see column 3 in Table 1) one
has the following reduced-error estimates for the longitudinal-to-transverse
scaling allometric exponents: 
\begin{equation}
\lambda ^{(buckl)}=0.87\pm 0.02\text{ and }\lambda ^{(bend)}=0.80\pm 0.03%
\text{, with }\lambda ^{(\exp )}=0.80\pm 0.02.  \eqnum{18}
\end{equation}
One can see that solely the elastic-bending criterium is validated. This
implies corroboration the bone evolution mechanism, which provides avoidance
of mechanical failure of mammalian bones caused by critical {\em elastic
bending} deformations induced by {\em maximum-area} muscle contractions
achieved in long bones during peak stresses.

From the physical point of view, the fact that the bending (but not
buckling) elastic deformations are crucial for mechanical failure of {\em %
long} rigid bones should be expected, under condition that the inequality $%
L_{is}\gg D_{is}$ (but not $L_{is}\gtrapprox D_{is}$) is fulfilled for
animals of arbitrary mass. Meanwhile, this fact was not corroborated in the
one-scale long-bone allometry, and we therefore report on the first
observation of the bending-critical-force bone-evolution mechanism, which is
suggested to be universal regardless of small and large mammals. Finally,
after McMahon, we have demonstrated how the scaling laws established in
mammalian allometry arise from a natural similarity of animals and how they
can be quite explicit on the evolution constraints through simple
geometrical and clear physical conceptions.

{\Large Acknowledgments}

The author is grateful to Per Christiansen and to Jayanth Banavar for their
scientific interest to the current research and to the Ph.D. student
Guilherme Garcia for helpful discussions of the biological literature.
Financial support by the CNPq is also acknowledged.\medskip

\newpage

\begin{center}
\begin{tabular}{|l||ll||lll||ll||ll|}
\hline\hline
Bone & dimen & sions & reduc & ed dim & ensions & bone & mass & muscle & area
\\ \hline
{\em LSR } data & $d_{i}$ & \multicolumn{1}{|l||}{$l_{i}$} & $\ l_{i}/d_{i}$
& \multicolumn{1}{|l}{$\lambda _{i}$} & \multicolumn{1}{|l||}{$%
b_{i}^{*}/d_{i}-2$} & $2d_{i}+l_{i}$ & \multicolumn{1}{|l||}{$\ \ b_{i}^{*}$}
& buckling & \multicolumn{1}{|l|}{bending} \\ \hline\hline
humerus & $.3816$ & \multicolumn{1}{|l||}{$.2996$} & $0.785$ & 
\multicolumn{1}{|l}{$0.763$} & \multicolumn{1}{|l||}{$0.804$} & $1.063$ & 
\multicolumn{1}{|l||}{$1.070$} & $0.927$ & \multicolumn{1}{|l|}{$0.838$} \\ 
\hline
radius & $.3868$ & \multicolumn{1}{|l||}{$.2995$} & $0.774$ & 
\multicolumn{1}{|l}{$0.753$} & \multicolumn{1}{|l||}{$0.802$} & $1.073$ & 
\multicolumn{1}{|l||}{$1.084$} & $0.948$ & \multicolumn{1}{|l|}{$0.850$} \\ 
\hline
femur & $.3548$ & \multicolumn{1}{|l||}{$.3014$} & $0.849$ & 
\multicolumn{1}{|l}{$0.843$} & \multicolumn{1}{|l||}{$0.988$} & $1.011$ & 
\multicolumn{1}{|l||}{$1.060$} & $0.816$ & \multicolumn{1}{|l|}{$0.714$} \\ 
\hline
tibia & $.3600$ & \multicolumn{1}{|l||}{$.2571$} & $0.714$ & 
\multicolumn{1}{|l}{$0.764$} & \multicolumn{1}{|l||}{$0.717$} & $0.977$ & 
\multicolumn{1}{|l||}{$0.978$} & $0.926$ & \multicolumn{1}{|l|}{$0.822$} \\ 
\hline\hline
Averaged & $.3708$ & \multicolumn{1}{|l||}{$.2894$} & $0.781$ & 
\multicolumn{1}{|l}{$0.781$} & \multicolumn{1}{|l||}{$0.828$} & $1.031$ & 
\multicolumn{1}{|l||}{$1.048$} & $0.904$ & \multicolumn{1}{|l|}{\bf 0.806}
\\ \hline
\end{tabular}

\begin{tabular}{|l||l|l||l|l|l||l|l||l|l|}
\hline
{\em RMA }data & .$d_{i}$ & \multicolumn{1}{|l||}{..$l_{i}$} & ..$\
l_{i}/d_{i}$ & $~......\lambda _{i}$ & \multicolumn{1}{|l||}{...$b_{i}{\sl /d%
}_{i}{\sl -}2$} & $2d_{i}+l_{i}$ & \multicolumn{1}{|l||}{$b_{i}$} & $\quad
a_{ci}^{(buckl)}$ & ~~ $a_{ci}^{(bend)}$ \\ \hline\hline
humerus & $.3860$ & \multicolumn{1}{|l||}{$.3109$} & $0.805$ & $0.784$ & 
\multicolumn{1}{|l||}{$0.806$} & $1.083$ & \multicolumn{1}{|l||}{$1.083$} & $%
0.947$ & $0.862$ \\ \hline
radius & $.4014$ & \multicolumn{1}{|l||}{$.3210$} & $0.800$ & $0.787$ & 
\multicolumn{1}{|l||}{$0.743$} & $1.124$ & \multicolumn{1}{|l||}{$1.101$} & $%
0.959$ & $0.874$ \\ \hline
femur & $.3599$ & \multicolumn{1}{|l||}{$.3089$} & $0.858$ & $0.864$ & 
\multicolumn{1}{|l||}{$0.976$} & $1.029$ & \multicolumn{1}{|l||}{$1.071$} & $%
0.850$ & $0.799$ \\ \hline
tibia & $.3654$ & \multicolumn{1}{|l||}{$.2767$} & $0.757$ & $0.804$ & 
\multicolumn{1}{|l||}{$0.731$} & $1.008$ & \multicolumn{1}{|l||}{$0.998$} & $%
0.851$ & $0.781$ \\ \hline\hline
Averaged & $.3782$ & \multicolumn{1}{|l||}{$.3044$} & $0.805$ & $0.810$ & 
\multicolumn{1}{|l||}{$0.814$} & $1.061$ & \multicolumn{1}{|l||}{$1.063$} & $%
0.901$ & {\bf 0.829} \\ \hline
\end{tabular}

.
\end{center}

.

Table 1. Testing of the mammalian long-bone similarity through the
elastic-buckling and elastic-bending criteria. Experimental data by
Christiansen on the mammalian dimension allometric exponents for $i$-bone
diameter exponent $d_{i}$, length exponent $l_{i}$, reduced dimension
exponent $\lambda _{i}$, and bone mass exponent $b_{i}$ obtained by the
least square regression ({\em LSR}) and the reduced major axis ({\em RMA})
regression methods. These are taken from Tables 2 in Refs.\cite{C99a},\cite
{C99b} and \cite{C02}, respectively. The {\em LSR} data on $b_{i}^{*}$ are
estimated here with the help of relation $b_{i}^{*}=r_{i}b_{i}$ , where $%
r_{i}$ (correlation coefficient) and $b_{i}$ are corresponding data obtained
by {\em RMA} regression. Predictions for the muscle-area critical exponents
are given with the help of Eq.(16). Bone-{\em averaged} magnitudes are found
as the mean values of the corresponding mammalian allometric exponents, {\it %
e.g.}, $d=\Sigma _{i=1}d_{i}/4$.

\newpage

\begin{center}
{\Large Figure Captures}
\end{center}

.

.

Fig. 1. Evolution of the cross-section area for muscle fibers with body mass
in the mammalian hindlimbs. {\em Points}: diamonds, circles, squares and
crosses are experimental data taken from Fig.3 in Ref.\cite{PS94} for,
respectively, gastrocnemius, plantaris, deep digital flexors and common
digital extensors. {\em Arrows} indicate the maximum muscle-area points
achieved for a given mass. {\em Solid line} corresponds to regression $%
A_{cm} $ of these points along with their nearest neighbors, with $%
A_{cm}=366*M^{0.82}$. {\em Dashes line} is given for the isotropic scenario
description with $A_{0}=29*M^{2/3}$.

.

.

Fig. 2. Mammalian bone-dimension diagram: bone diameter against bone length. 
{\em Points:} A'79, B'83, B'92 and C'99 correspond to the
overall-bone-averaged allometric data derived through the least square
regression method by Alexander {\it et al}., Biewener, Bertran \& Biewener,
and Christiansen and reported, respectively, in Refs.\cite{AJM79,B83,BB92}
and \cite{C99a}. {\em Crosses} correspond to the ESM\cite{Mc73} ($%
d_{0}^{(buckl)}=3/8$, $l_{0}^{(buckl)}=1/4$, $a_{0}=b_{0}=1$) and isometric
scenario ($d_{0}=l_{0}=1/3$, $a_{0}=2/3$, $b_{0}=1$) predictions; $a$-{\em %
lines} are due to the elastic-bucking and elastic-bending $a$-constraints
given in, respectively, Eqs. (9) and (13) and estimated for the case of the
critical-force exponent $a_{c}=0.82$ derived in Fig.1. The dashed areas
indicate the elastic-bucking and elastic-bending {\em criteria} given,
respectively, in Eqs.(10) and (14). These areas extend the corresponding $a$%
-lines by accounting of the $b$-constraint equations within the experimental
error for $a_{cm}^{(\exp )}=0.82\pm 0.01$ and $b^{(\exp )}=1.05\pm 0.05$
taken, respectively, from Fig.1 and Table 1. \newpage

\end{document}